\documentstyle[preprint,aps,epsfig]{revtex}

\begin{document}

\vskip 3cm

\centerline{\bf The LPM effect for EeV hadronic showers in ice:} 
\centerline{\bf  implications for radio detection of neutrinos}
\vskip 0.5cm
\centerline{J. Alvarez-Mu\~niz and E. Zas.}

\centerline{\it Departamento de F\'\i sica de Part\'\i culas, Universidade de
Santiago}

\centerline{\it E-15706 Santiago de Compostela, Spain}

\begin{abstract}
We study the longitudinal development of hadronic showers in ice for energies 
up to 100 EeV and its implications for radio and optical Cherenkov emission. 
A small fraction of showers induced by primaries of energy above 1~EeV 
is shown to display the characteristic elongation associated to the Landau Pomeranchuck 
Migdal (LPM) effect. The rest look like ordinary showers because the highest energy
$\pi^0$'s interact 
instead of decaying in two photons. 
The LPM effect observed in this fraction of the showers is mainly due to 
$\eta$ and $\eta '$ production and decay. We give parameterizations for the total 
and excess charge tracklengths and for the amplitude and angular spread of the 
electric field spectrum in the Cherenkov direction.
Implications for neutrino detection are briefly addressed.
\end{abstract}

\vskip 1cm
{\bf PACS number(s):} 96.40.Pq; 29.40.-n; 96.40.Kk; 96.40.Tv


\newpage
\section{Introduction}
The detection of low energy neutrinos (MeV-GeV) from the Sun and from cosmic 
ray interactions in the atmosphere is now routine in underground detectors which 
we hope will reveal new aspects of the neutrino sector in the Standard Model
\cite{concha}. 
On the other hand the wealth of information obtained with the detection of 
few MeV neutrinos from supernova SN1987A showed that the enormous potential of 
neutrino astronomy could be reality in the near future. Several experiments 
are being constructed at present to detect neutrinos at higher energies, at which   
the atmospheric neutrino flux is expected to be sufficiently low to obtain  
new information from extraterrestrial sources. They were conceived 
to measure the Cherenkov light emitted by upcoming muons produced 
in charged current neutrino interactions that cross the Earth. The long muon ranges allow 
an effective volume sufficiently large to answer important questions, for 
example on the 
intriguing acceleration mechanisms in Active Galactic Nuclei (AGN) and Gamma 
Ray Bursts (GRB)\cite{physreps}.  

It is becoming increasingly apparent that the highest energy 
neutrinos will play an important role. The identification of the highest energy 
gamma ray sources 
with Blazars, which are interpretated as AGN with jets pointing in our 
direction, has shifted theorists to model acceleration in the jets. These jets  
are believed to be particle flows with relativistic bulk motion and high 
associated Lorentz factors which naturally make the emission highly anisotropic 
and boost it to very high energies\cite{padovani}. 
It is also believed that the infrared background prevents the TeV gamma ray 
emission from reaching the Earth except for a few nearby sources\cite{stecker}. 
The possibility is open that particles are accelerated to much higher 
energies getting absorbed in the different backgrounds. Similarly for GRB's 
there are acceleration models reaching up to energies much greater than observed. 
Both phenomena are candidates for the intriguing origin of the 
highest energy cosmic rays\cite{biermann,waxman}. 
If this is the case, large fluxes of neutrinos can be expected up to close to 
the highest energies the hadrons are accelerated. 
In any case EeV neutrinos must 
exist because of the interactions of cosmic rays with the microwave photons 
and their search can answer important 
issues with respect to these objects or other potential neutrino sources 
as well as to the origin of the highest energy cosmic rays. 

EeV neutrinos have a high enough cross section that   
prevents them from traversing the Earth without interacting.
These high energy neutrinos produce extremely high 
energy showers whose Cherenkov light may also allow efficient detection 
in the underground detectors in project or construction. Alternative detection mechanisms become favourable for extremely 
energetic showers.  
The radio technique relies on the coherent emission of Cherenkov light in 
radio waves by all shower particles when the wavelength of the emission is 
larger than the shower dimensions\cite{askaryan}. 
Ice is being considered as a good 
possibility because of the long attenuation length for radio waves\cite{markov}. 
Similarly the proposed acoustic alternative 
relies on detecting the coherent sound wave which is produced by the shower in  
ocean water\cite{learned}. The coherence  
condition makes the signals scale as the square of the incident particle energy 
and this is the reason why these methods become most promising at the highest 
energies. Depending on the actual fluxes the construction of large enough detectors 
sensitive to the expected fluxes may be cost effective provided these alternatives 
are succesfully implemented\cite{jelley}. 
Several estimates of the neutrino expected event rates
in these detectors have been made\cite{ralston,price,provorov}. 
An interesting possibility is the detection of the radio pulses 
from cosmic ray and neutrino interactions on the Moon surface
using radiotelescopes on Earth \cite{jaime}. EeV shower behavior in dense media is important
for EeV neutrino detection, particularly in water (or ice) as chose for most of the proposed
detectors.

Above PeV energies the cross sections for pair production and 
bremsstrahlung get suppressed because the characteristic length of the interaction becomes 
larger than atomic spacing and collective effects 
of the atomic fields have to be considered.
This is the Landau Pomeranchuck Migdal (LPM) effect \cite{LPM}. Electromagnetic EeV showers 
have been shown to have huge elongations because of the 
LPM effect \cite{misaki,stanev,alvarez} and hadronic showers are expected to behave
similarly if they contain PeV photons or electrons.

Here we study the development of EeV hadronic showers in 
water concentrating on the relevant parameters involved in the emission of 
Cherenkov light and coherent radio pulses. We 
will use many results presented in the study of electromagnetic showers 
from the same perspective as described in ref \cite{alvarez}. 
As it was the case for 
electromagnetic showers, the full simulation of Cherenkov light 
and in particular coherent effects between different particles is not practical 
because of the high energies involved. We choose to approximate shower 
development by a combination of Monte Carlo methods and parameterizations of
showers of lower energies. The main characteristics of the radio pulse 
emission from these showers at the Cherenkov angle can be deduced from the 
longitudinal shower development so we will use one dimensional approximations
which will only be valid for emission in this direction. Although more detailed 
calculations will be necessary, many of the most important aspects of the 
possibilities of detecting hadronic showers as induced 
by neutrino interactions can be answered with the approximations made here.  
Many of the qualitative conclusions obtained in this work
can be easily extended to other media.

\section{Hadronic showers.}

Shower development has been studied in air up to $10^{21}$~eV 
because all high energy cosmic ray experiments look for them. At lower energies
($<1$~TeV) they have 
also been studied in dense media for accelerator experiments. 
The study of purely electromagnetic showers and LPM implications, 
has been done in water for energies up to 10~EeV \cite{alvarez}. 
Such showers are produced at the lepton 
vertex in charged current Deep Inelastic Scattering (DIS) electron neutrino interactions. 
In these interactions the average energy transfer to the lepton  
is predicted to be large, about $75\%$\cite{sarcevic}. 

Hadronic showers produced in both charged and neutral current 
DIS neutrino interactions are started by hadrons in the fragmentation of the nuclear 
debris. At 
high energies the multiplicity is expected to be about half of that in 
$e^+e^-$ collisions for a center of mass energy squared equal to minus the square of the
4-momentum transfer ($Q^2=-q^\mu q_\mu$) in the DIS process \cite{DISinclusive}. 
Practically all models for neutrino production involve the decay 
of pions which naturally predict more muon 
flavor neutrinos from naive channel counting. The ratio of muon to electron 
neutrino production may be actually larger than expected 
in some environments because of muon 
synchrotron losses as has been recently stressed\cite{rachen}. All these 
factors may well compensate the fact that in DIS interactions only one 
fourth of the neutrino energy is typically transferred to the hadron vertex. 
Multiple hadrons are also produced in $68\%$ of the $W$-boson decay in the 
resonant electron antineutrino scattering with atomic electrons. 
Hadronic showers will also play an important role in tau neutrino detection 
through the proposed double bang events \cite{pakvasa}. 

For EeV neutrinos the fragmented particles, mostly pions, 
will mantain the direction of the primary neutrino in the lab frame 
because their average transverse momentum is expected to be in the few hundred 
MeV range. These particles continue to interact with nucleons and they still 
have sufficient energy so that after three or four generations the cummulative angular deviations are still very small. The hadronic shower can be thought as having a hard penetrating central core containing mostly pions which feeds electromagnetic subshowers fundamentally through $\pi^0$ 
decay in two photons. Because the medium is dense, charged pions are expected 
to interact before decaying. Assuming energy equipartition between all flavor pions, a fraction of 1/3 would go into electromagnetic subshowers every time there is an interaction. If all interactions occur simultaneously every interaction length (to make things simple) it is easy to find that after three generations $80\%$ of the energy is in electromagnetic subshowers. The shower will have a markedly electromagnetic character because of the 
high energy involved and the medium density which prevents decays of most 
of the charged pions. 
The lateral distribution should be related to a sum of 
distributions of electromagnetic showers started at different depths. 
Near the core the showers are thus
younger, dominated by small electromagnetic showers initiated deeper, while away from the shower axis they are due to early electromagnetic subshowers of 
higher energies. 

Continuing with this simple model, we can consider that after a neutrino 
interaction with pion multiplicity $N$, the pion energy is roughly 
the energy transferred to the hadron $yE$ divided by $N$. 
The LPM effect is known to increase the electron and photon 
interaction lengths above some threshold energy $\sim E_{LPM}$ (2~PeV for ice) 
\footnote{Corresponds to the definition in Ref.~\cite{stanev}.}
\cite{stanev,misaki,klein}. 
However the  
longitudinal profile of electromagnetic showers in ice around shower maximum 
starts to be affected typically at  
primary energies above $E=20~$PeV\cite{alvarez}. 
For a 1 EeV energy transfer to the hadron a multiplicity of order 15 can be expected
in the $\nu$ interaction and the photons from the $\pi^0$ decays would roughly have 30~PeV. 
Moreover most of the electromagnetic 
energy in the shower is produced after the first generation. 
As a result little deviations from 
typical shower development are expected below 1~EeV because of the LPM effect. 

As the primary energy rises so will the energy of the first generation $\pi^0$'s. 
However they do not necesarily produce high energy photons. 
Neutral pions have a decay length $c \tau = 2.5~10^{-6}~$cm and an interaction depth 
of about $130~$gcm$^{-2}$ so that in the competition between decay and interaction 
the latter dominates for $\pi^0$ energies   
above about 6.7~PeV. As a result the showers can be expected to show LPM 
effects in a mitigated form. Even some 100 EeV showers are actually shown below 
to display small LPM elongations. EeV hadronic showers 
can look more like rescaled versions of lower energy  
showers while EeV electromagnetic showers are greatly distorted and elongated 
because of the LPM effect. 

\section{Simulations.}

A complete simulation of the radio pulse should monitor particles down 
to the MeV range requiring unreasonably large times for EeV energies\cite{Zas}. 
To investigate EeV hadronic shower characteristics we have developed a fast 
hybrid Monte Carlo which simulates 1 dimensional 
showers down to some crossover energy, at which the subshower 
produced is taken from a tested parameterization. 
For electromagnetic subshowers we use the Greisen parameterization which was 
shown in \cite{Zas} to be valid up to energies of 100~TeV. 
We have calculated our own parameterizations for hadronic 
subshowers inititated by protons, pions and kaons. 
The code has been designed 
to calculate quantities such as the projected tracklength (onto shower axis) 
and the weighted projected tracklength (excess tracklength due to electrons). 
These quantities are respectively known 
to be the relevant parameters for optical and radio 
Cherenkov emission\cite{Zas}. We have used parameterizations of these tracklengths for electromagnetic showers using the results obtained in \cite{Zas}. 
Intermediate results will be presented elsewhere. 

The shower simulation part is based on UNICAS\cite{Gaisser} which has been 
adapted to run in a homogeneous water medium with density $0.924~$gcm$^{-3}$. 
The code has also been modifyed to 
take into account interactions of $\pi^0$'s and other shortlived resonances as 
well as corrections to bremsstrahlung and pair production accounting for the 
Landau-Pomeranchuk-Migdal effect. 
For the lower energy showers we have performed full simulations. 
Above 100 TeV we use the hadronic code SIBYLL\cite{SIBYLL} together with the 
parameterizations.
SIBYLL includes ''minijet events'' which are 
responsible for the cross section increase with energy\cite{halzen} and soft 
collisons based on the Dual Parton Model (DPM) \cite{capella}. 

We have made extensive independent checks of the codes used. For example 
full 1-D simulations (for energies below 100 TeV) are well within the 
uncertainties associated to the different hadronic models compared with 
CORSIKA in \cite{Knapp}. We have also checked the 1-D version of UNICAS in 
ice comparing photon showers with those obtained in the 3-D 
electromagnetic code developed specifically for calculating radio pulses in 
ice\cite{Zas}. 
We have checked the hybrid approach both comparing it to  full 
simulations and 
ensuring that all hybrid codes are stable under changes of  
the energy crossover. 

\section{Results.}

We have simulated showers initiated by protons, and most of the results 
presented here correspond to such showers. 
Little difference is observed between proton and pion initiated showers 
of the same primary energy. We have explored the energy range 
from 10 GeV to 100 EeV simulating large numbers of proton showers 
(over 20,000 showers above 1 EeV) with the described codes. 
The lower energy showers are necessary to make parameterizations which 
allow the simulation of the highest energy ones. 

We firstly show the results of the fraction of energy going into 
electromagnetic showers in Fig. 1. 
Here the fraction is seen to rise with 
energy according to the simplified model above, until the effect of neutral 
pion interactions becomes important. In the total tracklength calculation, 
relevant for the Cherenkov light output, the electromagnetic contribution 
accurately scales with the electromagnetic energy in the shower as shown in 
Fig. 1. Other charged particles contribute some percentage 
which decreases with energy (15 $\%$ at most) as can be expected from the 
electromagnetic energy fraction in Fig. 1. The coherent radio pulse emission 
scales with the difference between positive and negative charge 
tracklengths, the weighted tracklength, to which particles other than 
electrons and positrons contribute negligibly. We also display 
''shower length'' in Fig. 1, defined in \cite{alvarez} 
as the length along which shower size exceeds $70\%$ of its maximum. 
The angular width of the Cherenkov radiopulse emitted at 
the Cherenkov angle is approximately inversely proportional to this parameter.

Hadronic and electromagnetic showers have well known differences which have 
been studied at high energies only in air, mainly  
hadronic showers have larger attenuation length\footnote{Attenuation
length is defined as the length over which 
the energy drops by a factor $e$ in the region well after shower 
maximum and it is proportional to the particle interaction length.},
reach deeper from first interaction point, have harder lateral distributions and 
carry a significant fraction of their energy in muons and neutrinos. 
As a result they have slightly less particles at maximum than electromagnetic 
showers of the same primary energy. These differences can be seen in Fig.~2 
where proton showers are compared
to  pure electromagnetic showers of energies $10$~TeV, $1$ and $100$~PeV. 
As long as the 
LPM effect is not significant, the showers "scale" with primary 
energy in much the same way as electromagnetic showers, what incidentally, 
simplifies their parameterizations. 

Below 1~EeV the longitudinal profile of hadronic showers has this scaling behavior,
characterisitc of showers that are not affected by the LPM effect. 
This is not surprising, 
the average energy of the pions 
is well below the primary interaction energy because of the large 
multiplicities involved \cite{multiplicities} and only electromagnetic 
showers of energy above $\sim 20~$PeV show "scaling deviations"\cite{alvarez}. 
In table I we give the mean energy of the $\pi^0$'s 
as a function of primary energy as obtained in the hadronic interaction 
generator. In Fig. 3 we compare the energy distributions of all the first
generation $\pi^0$'s 
in proton-ice interactions at different energies to those that actually decay
into photons. 
Only about 10$\%$ of the $\pi^0$'s are expected to produce photons of 
energy above $20~$PeV because most of them interact. 
Showers are not elongated despite being 
induced by primaries exceeding $E_{LPM}$ by several orders of magnitude. 
In Fig. 1 we also display shower length 
neglecting neutral pion interactions to further illustrate this point. 

In this respect the role of resonances with shorter lifetimes than the $\pi^0$ 
can be more important if they have decay channels involving electrons or 
photons. 
If these resonances are produced in early interactions, high energy photons (or electrons)
produce subshowers with characteristic LPM elongations. 
The extent of this effect is a probabilistic issue 
and depends on detailed aspects of the hadronic interaction model. 
At the highest energies, experimental evidence is non-existant and our results 
have to be taken with caution. 
They indicate that these aspects are important for shower development 
and it may be that, if these showers are ever observed, they would 
provide experimental information on such channels. 
In particular we 
find that the $\eta$ and the $\eta '$ contribute most and this is 
just because they are very short lived, they are produced relatively often and 
they have considerable branchings into photons, $\sim 40\%$ ($\sim 65\%$) for 
$\eta$ ($\eta'$\footnote{Including the $\eta' \rightarrow \eta~\pi^+\pi^-$ decay chain.}). 
 
We find that a proportion of showers above 1~EeV have deep tails 
characteristic of LPM showers (see Fig. 4). 
We also note that the energy content of these tails (proportional to shower development
integral) is typically of order 1$\%$. All of these can be quantitatively addressed
with the aid of Fig. 5, representing the "leading photon" distribution as obtained in SYBILL, that is the
differential probability that the highest energy photon in a shower has a fraction energy
($x=E_{\gamma}/E_0$) in the range ($x$, $x+$d$x$). 
The median is at about  $2\%$ and not very sensitive to primary energy, 
hence about 50$\%$ of the hadronic showers having energy above $50~E_{t}$ 
are expected to have a photon of energy $>E_{t}$. If $E_{t}$ is chosen say above 100~PeV, 
the electromagnetic subshower generated by the leading photon will have a 
long LPM tail. The range of leading photon energy fractions is between 
$0.1\%$ and $30\%$ in $99.9\%$ of the showers and one can easily deduce that 
it is practically impossible to 
have a shower without an LPM tail if the primary energy is 
above 100~EeV and that hadronic showers of energy below 250~PeV will hardly have LPM effects.

\section{Cherenkov radiation from hadronic showers.}

Cherenkov light is emitted as charged particles travel 
through a medium at a  speed greater than that of light. 
While the electric field amplitude 
is proportional to particle tracklength, in the optical 
region the emission is incoherent and the power output 
scales with the total tracklength\cite{Frank}. For wavelengths exceeding  
the shower dimensions Cherenkov light is emitted coherently by all 
shower particles, opposite charges 
contribute with opposite phases and the electric 
field amplitude is proportional to the weighted tracklength\cite{allan}. The excess 
electrons in electromagnetic showers are caused by low energy interactions 
with atomic electrons \cite{Zas,allan}. 
The matter antimatter symmetry in the cross sections 
is broken by the  absence of target positrons. The 
dominant process is 
Compton scattering with atomic electrons, but electron positron 
annihilation, M\"oller and Bhabha scattering also contribute to an overall 
$21\%$ excess when expresssed in terms of the total tracklength \cite{Zas}. 
Many of the results obtained in the context of radio emission 
from electromagnetic showers \cite{alvarez,Zas} can be applied to   
our hadronic shower simulation results.

There is a smooth transition from the incoherent 
to the coherent Cherenkov regions which has been 
described in \cite{alvarez,Zas}. 
It occurs in ice at $\sim$1 GHz for radiation emitted in the Cherenkov angle 
direction and at lower frequencies away from this direction. 
The spectrum rises linearly with 
frequency until interference effects from different shower regions start to 
take place. This happens at a maximum 
frequency $\nu_{max}$. For pulses observed at the 
Cherenkov angle the interference is
governed by the lateral distribution, the narrower the lateral spread
the bigger the maximum frequency. We expect it not to be too different 
from electromagnetic showers, and we take $\nu_{max}$ from them. As a 
first approximation this 
is justified because most energy in hadronic showers is electromagnetic, and 
consists of much lower energy subshowers.    

The electric field amplitude in the Cherenkov direction can 
be approximated quite accurately as a 1-D Fourier transform of the longitudinal 
charge excess distribution. The half width of the central peak is 
inversely proportional to shower length as defined above 
and shown in Fig. 1. 
This has been used to describe Cherenkov emission 
from EeV electromagnetic showers in ice\cite{alvarez}. 
Since the 
highest output signal is concentrated in the Cherenkov peak, 
its size and angular spread are the most relevant aspects of such radiation, 
particularly 
for establishing the detection capabilities of future detectors. 
These properties are calculable in this approach and can be used for detector 
design while more detailed simulations are implemented.   

The change in shower length with 
energy is very small, $\sim 30\%$ 
between 100~TeV and 10~EeV as shown in Fig. 1. As a result the 
angular distribution of the Fourier transform of the radiopulse is not 
going to become narrower for higher energies contrary to 
electromagnetic showers \cite{alvarez}. For those showers that do display 
LPM effects the narrowing can be observed in a rather convoluted diffraction 
pattern, see Fig.~6; the Cherenkov radiopulse being 
coherent contains all the information on the excess charge distribution. 
The lateral distributions of the excess charge also affects the full 
diffraction pattern behavior in other directions 
but this is not addressed in this work. 

From our shower results it is straigthforward to obtain a parameterization 
for the total tracklength $t$ from which the total Cherenkov light in the optical
band can be obtained:

\begin{equation}
f(\epsilon)=-1.27~10^{-2}-4.76~10^{-2}~({\epsilon +3})-
2.07~10^{-3}~({\epsilon +3})^2+
0.52\sqrt{\epsilon +3 }
\end{equation}
\begin{equation}
\epsilon=log_{10}~{E_0\over {\rm 1~TeV}}
\end{equation}
\begin{equation}
t=6.25~f(\epsilon)~{E_0\over{\rm 1~GeV}}~{\rm m}
\end{equation}
The weighted projected tracklength is $\sim 0.21~t$ and the normalization of the 
radiopulse electric field spectrum at the Cherenkov peak $\vec E(\omega, R,\theta_C)$
scales with it:

\begin{equation}
R \vert \vec E(\omega, R,\theta_C) \vert=
1.1 \times 10^{-7}~{\rm f(z)}~{E_0 \over 1~{\rm TeV}}~{\nu \over \nu_0}~
{1 \over 1 + 0.4 ({\nu \over \nu_0})^2} \; {\rm V~MHz}^{-1}
\end{equation}
where $\nu$ is the observation frequency and $\nu_0\approx 500~{\rm MHz}$.

Lastly the angular spread of the electric field is inversely proportional to shower
length:

\begin{equation}
E(\omega,R, \theta)=E(\omega,R, \theta_{C})~
\rm e ^ {- {ln 2} \left[ { \theta -\theta_{C} \over \Delta \theta}
\right]^2 }
\end{equation}
\begin{equation}
\Delta \theta \simeq\cases{
1^{\rm o} ~ {\nu_0 \over \nu} ~
(2.07-0.33~{\epsilon}+7.5~10^{-2}~{\epsilon}^2) ~~{\rm for}~~1~{\rm TeV}<E_0<100~{\rm TeV}~\cr
1^{\rm o} ~ {\nu_0 \over \nu} ~
(1.74-1.21~10^{-2}~{\epsilon})~~{\rm for}~~100~{\rm TeV}<E_0<100~{\rm PeV}~~\cr
1^{\rm o} ~ {\nu_0 \over \nu} ~
(4.23-0.785~{\epsilon}+5.5~10^{-2}~{\epsilon}^2)~~{\rm for}~~100~{\rm PeV}<E_0<10~{\rm EeV}~} 
\end{equation}

\section{Discussion.} 

We have shown that EeV hadronic showers produced in neutrino interactions 
are very different from electromagnetic showers. 
The discussion above applies to showers initiated by protons or pions 
of a given primary energy. A 100~EeV neutrino  
transferres 25$\%$ of its energy to the hadronic debris with typical 
multiplicity 17 and only the leading particle keeps 
a significant fraction $1-K$ of the primary neutrino energy, where $K$ is the mean inelasticiy. 
As a result practically all of the electromagnetic subshowers will have 
energies far removed from the neutrino energy. We can expect 
the hadronic showers induced by neutrinos up to $10^3$~EeV 
to have a quite ordinary longitudinal development curves around shower maximum. 
They will however 
differ considerably from the electromagnetic showers produced in charged 
current electron neutrino interactions which display a typical fluctuating 
nature, reaching a great deal further and having many 
less particles at shower maximum\cite{alvarez}. 

It should be in principle possible to distinguish 
them comparing the shower profiles if they occur inside optical Cherenkov 
detectors such as those now in construction. 
For radiodetection the differences are converted to differences in the 
angular width around the Cherenkov direction. This effect favours the 
detection of hadronic induced pulses because the antenna spacing would 
not have to be reduced. It is
not clear that at EeV energies the electron
neutrino contributions for the electromagnetic pulses dominate. The answer
depends crucially on the relative electron to muon neutrino fluxes, the
fraction of energy transferred to the nucleons in DIS 
interactions and the relative spacing
between antennas.
With an array 
of antennas sufficiently dense it should be possible to recognize the 
much broader pulses for EeV hadronic showers. This is very interesting since 
the relative rates of each type can be related to the ratio of electron to 
muon neutrinos. This could be used to test neutrino mixing parameters  
models in regions of parameter space unaccesible to other experiments 
and similar to those suggested in Ref.~\cite{halzentau}.  

\vskip 0.5 cm
{\bf Acknowledgements:} We thank T. Stanev for making   
UNICAS available to us and for many discussions 
and suggestions on this work, we also thank
R.A. V\'azquez for useful comments and for carefully
reading the manuscript and L.G.~Dedenko, F. Halzen, J. Knapp, 
and P.~Lipari for fruitful discussions. This work was supported in part
by CICYT (AEN96-1673) and by Xunta de Galicia (XUGA-20604A96). One of 
the authors J. A. thanks the Xunta de Galicia for financial support.

\newpage

\begin{center}
FIGURE CAPTIONS
\end{center}

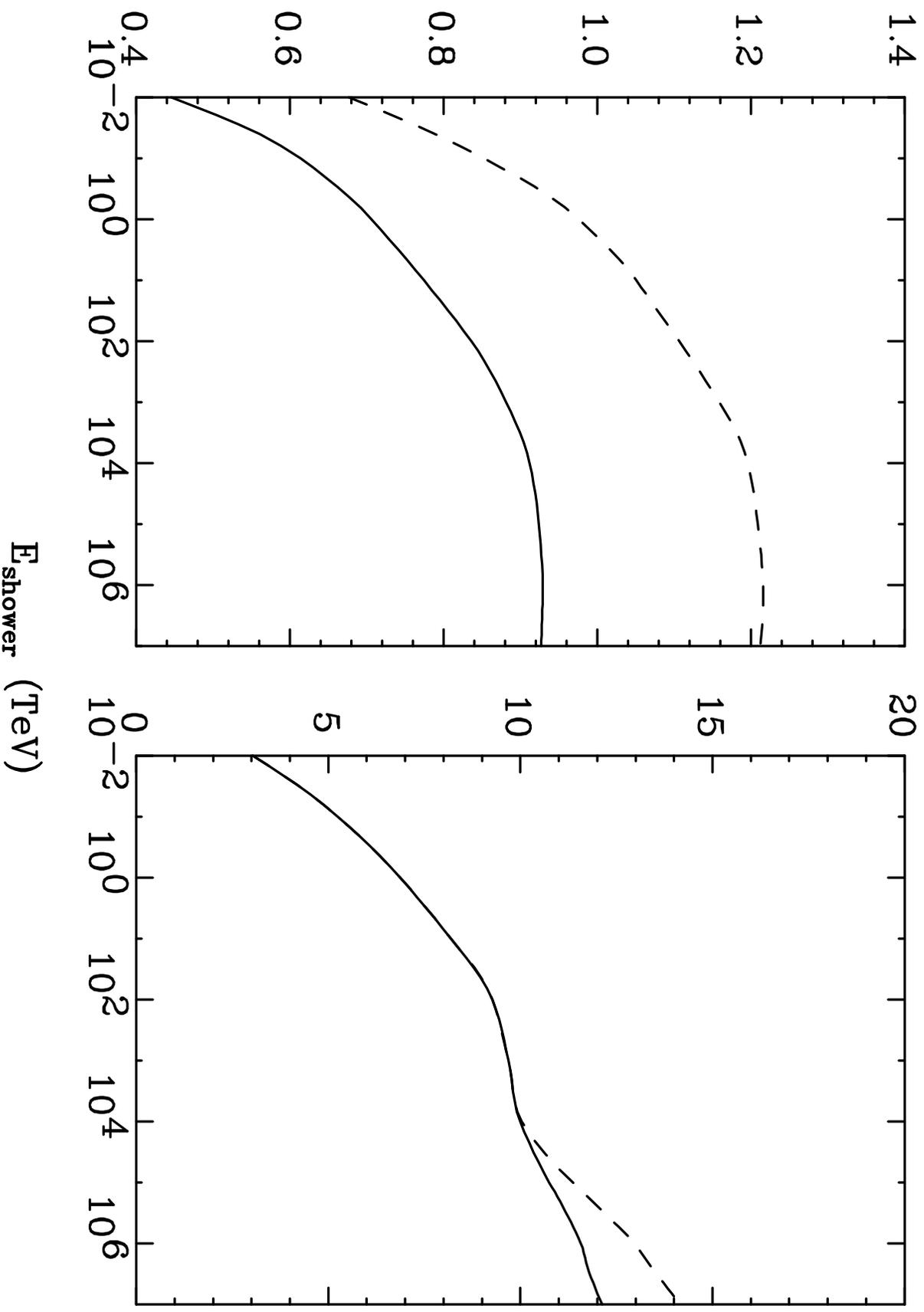
{\bf Figure 1:} Left: Electromagnetic energy fraction as a 
function of primary energy (solid)
and ratio of the weighted projected tracklength and shower energy in units
of km~TeV$^{-1}$.
Right: Shower length (in radiation lengths and defined in text) in
a proton shower as a function of primary energy.
The solid (dashed) line represents shower length
when competition between interaction and decay for $\pi^0$'s 
is (is not) performed.

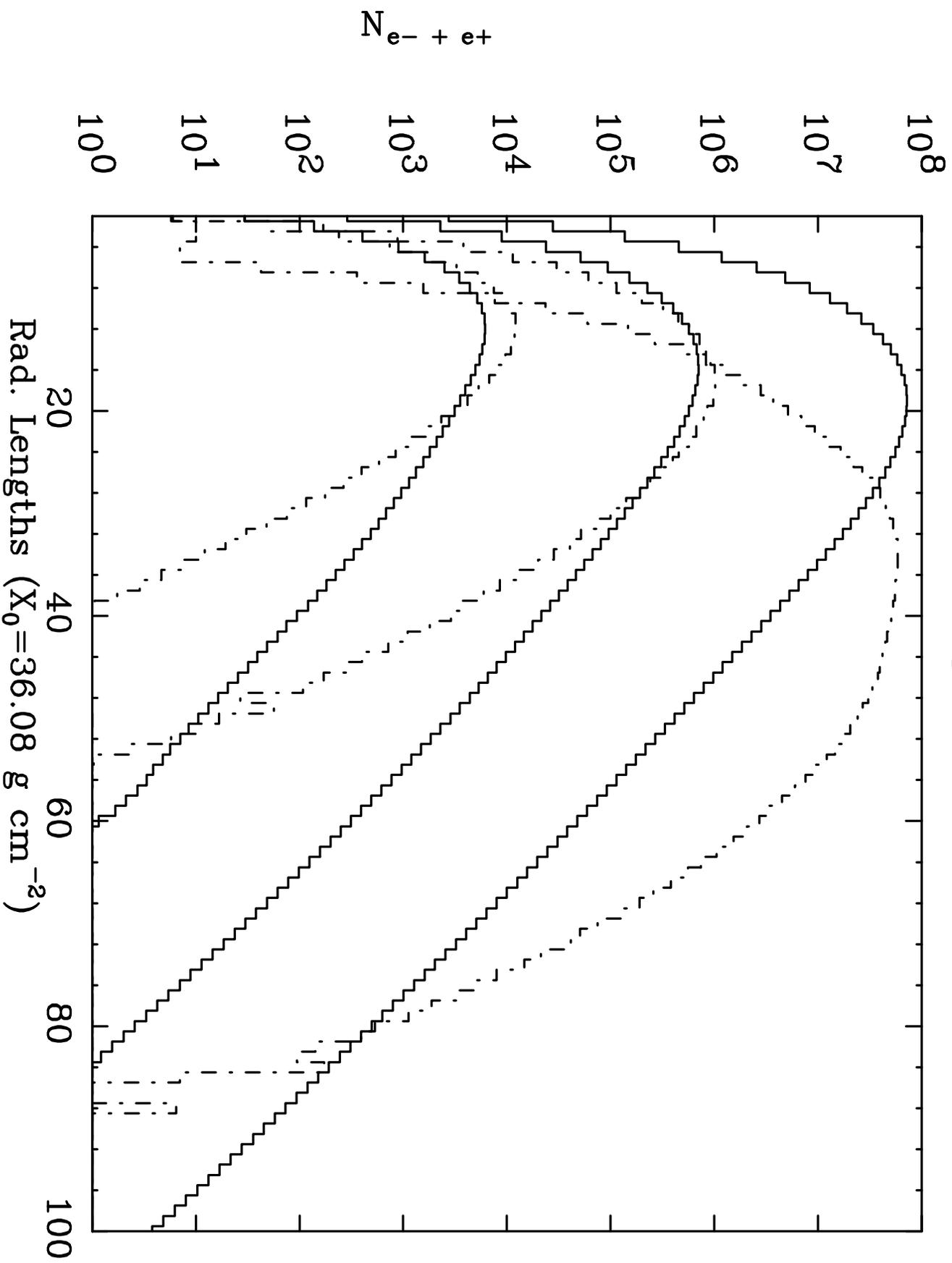
{\bf Figure 2:} Shower development of
a proton (solid) and electromagnetic (dashed) showers
of primary energies 10 TeV, 1 PeV and 100 PeV.

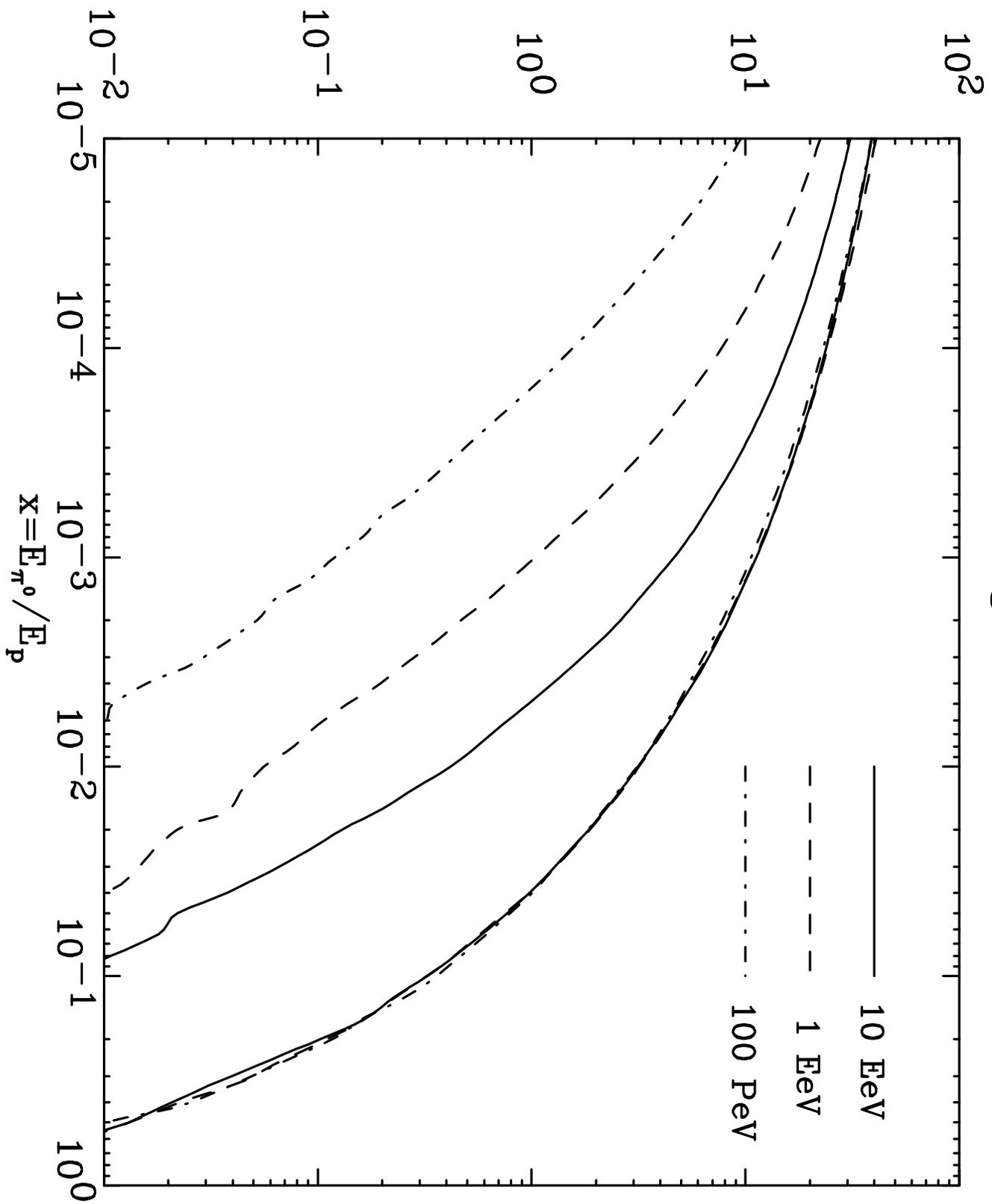
{\bf Figure 3:} Energy distribution of $\pi^0$'s in 10 EeV, 1 EeV 
and 100 PeV proton-ice
 interactions showing
the average number of $\pi^0$'s with energy greater than $x E_{\pi}$
($10^4$ events). The three overlapping upper curves 
correspond to all the  $\pi^0$'s
produced while the three lower curves count only those $\pi^0$'s that
decay. 

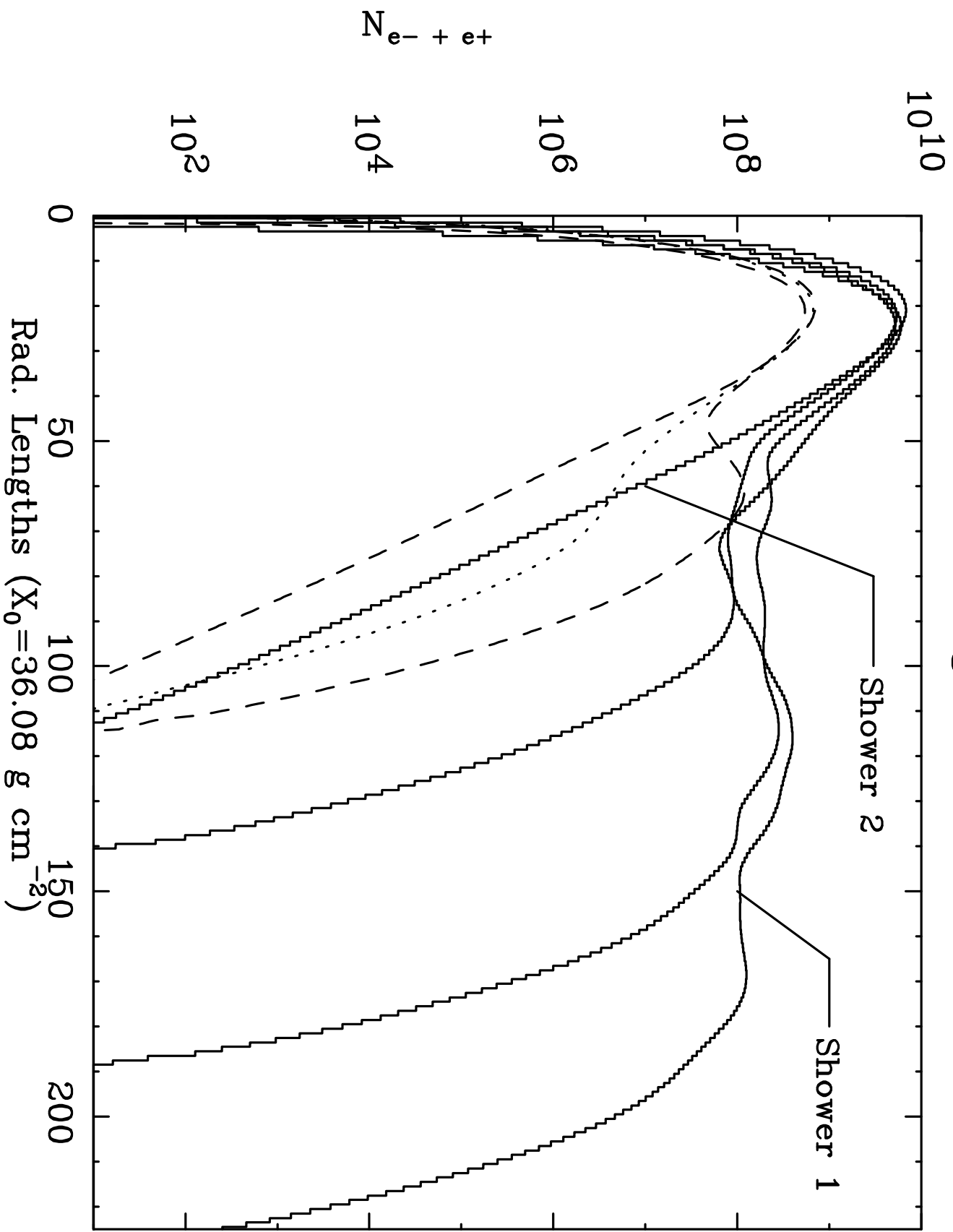
{\bf Figure 4:} Shower development of individual showers
of energies 10 EeV (solid histograms) and 1 EeV (dashed lines). For 1 EeV we also show the average over 100 showers (dotted line). Labels refer
to the individual showers taken to obtain Fig.~6.

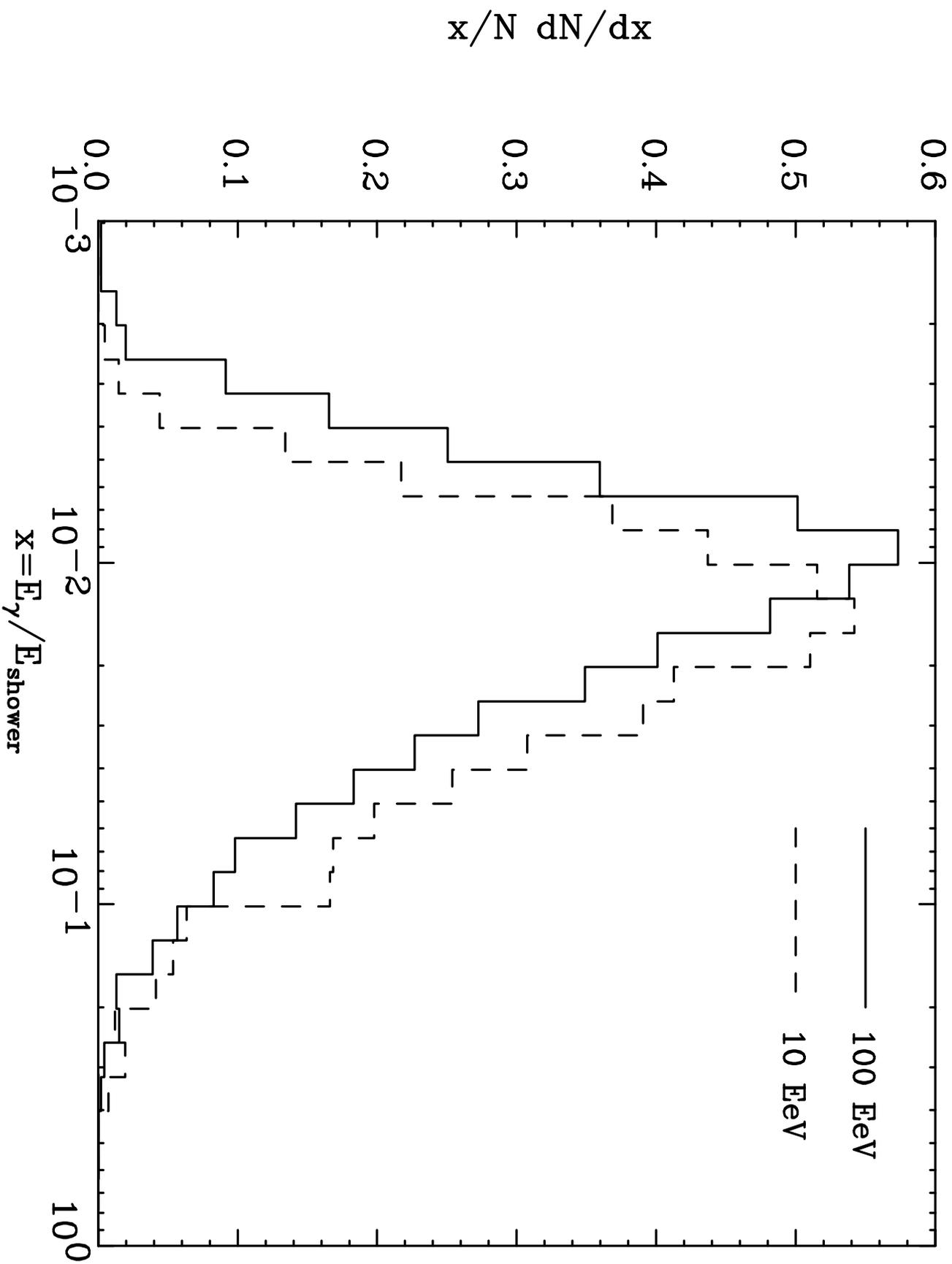
{\bf Figure 5:} Leading photon distributions for 10 EeV and 
100 EeV hadronic showers.
The area under this graph, as shown, is proportional to the probability.

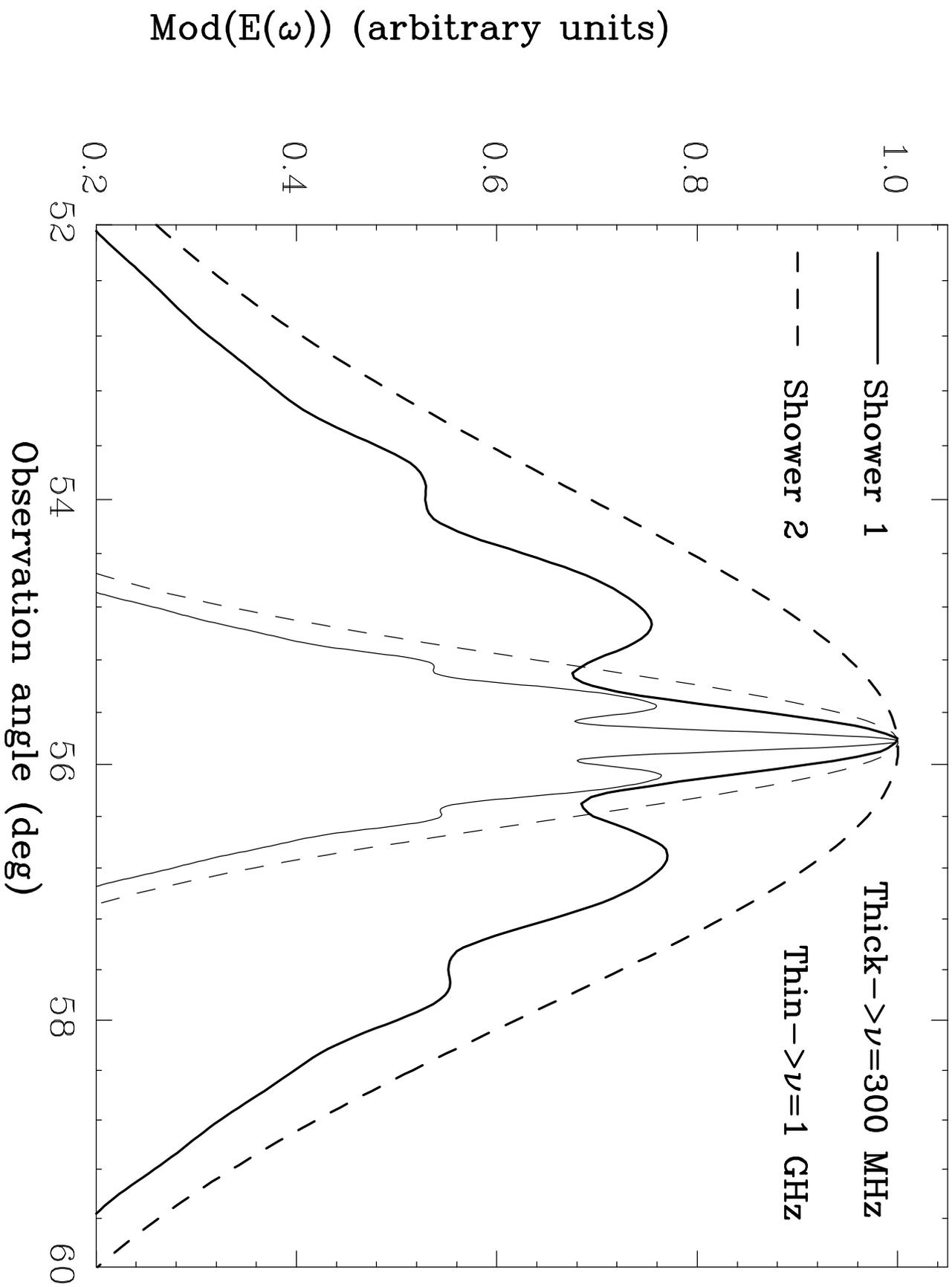
{\bf Figure 6:} Radiopulse distribution around 
the Cherenkov direction for the two
showers labelled in Fig.~4, illustrating the effect of the LPM tail on the
radiopulse angular distribution.

\begin{center}
TABLES
\end{center}

\begin{center}
\begin{tabular}{|c||c|c|c|c|}\hline
$E_0$ (TeV) &$10^4$ &$10^5$ &$10^6$ &$10^7$ \\ \hline\hline
$<E_{\pi^0}>$ (TeV) &40 &300 &$2~10^3$ &$1.7~10^4$ \\ \hline
\end{tabular}
\end{center}

{\bf Table I:} Average neutral pion energy in proton proton collisions 
for different projectile energies in the lab frame as produced by SIBYLL
hadronic generator.

\end{document}